\documentclass{PoS}
\usepackage{epsfig}
\usepackage{graphicx}
\usepackage{amsmath,amssymb}
\usepackage{colortbl}
\usepackage{color}
\usepackage{times}
\usepackage{slashed}
\usepackage{multirow}

\newcommand{\beq}{\begin{equation}}
\newcommand{\eeq}{\end{equation}}
\newcommand{\bea}{\begin{eqnarray}}
\newcommand{\eea}{\end{eqnarray}}

\newcommand{\eps}{\epsilon}
\newcommand{\ord}[1]{{\cal{O}}( #1 )}
\newcommand{\B}{{\bf B}}
\newcommand{\Bdag}{{\bf B^\dagger}}

\title{Baryon masses and axial charges  in $\mathbf {1/N_c}$ - ChPT}

\ShortTitle{Baryon masses and axial charges in $1/N_c$-ChPT}
\author{\speaker{\'Alvaro Calle Cord\'on}%
         \thanks{
We thank Dru Renner for discussions and comments on this article. 
Work supported by DOE Contract No. DE-AC05-06OR23177 under which JSA operates the
Thomas Jefferson National Accelerator Facility and by the National Science
Foundation   through grant PHY-0855789 (JLG). }\\
        Thomas Jefferson National Accelerator Facility, Newport News, VA 23606, USA \\
        E-mail: \email{cordon@jlab.org}}

\author{Jos\'e L. Goity$^{\dagger}$\\
Department of Physics, Hampton University, Hampton, VA 23668, USA.\\
Thomas Jefferson National Accelerator Facility, Newport News, VA 23606, USA \\
        E-mail: \email{goity@jlab.org}}

\abstract{An analysis  is presented  of the ground state baryon's ($N$ and $\Delta$) masses and axial couplings in the combined    $1/N_c$ and chiral expansions.  Renormalization and alternative power counting schemes are discussed to one loop (NNLO).
An application is made to lattice QCD results for non-strange baryon masses and the nucleon axial coupling as functions of the pion mass. This work represents a  brief  summary of a more extensive analysis to be presented elsewhere.}

\FullConference{Sixth International Conference on Quarks and Nuclear Physics\\
		 April 16-20, 2012\\
		 Ecole Polytechnique,  Palaiseau,  Paris}

\begin{document}

\vspace*{-1cm}

\section{Introduction}\vspace*{-2mm}
%

%
It has been known for a long time that the low energy expansion in the baryon sector  is  significantly improved if the spin 3/2 baryons are included as explicit degrees of freedom \cite{Jenkins:1990jv}.  It was soon realized that the reason behind the improvement is grounded in the $1/N_c$ expansion of QCD~\cite{'tHooft:1973jz}. In particular, the large $N_c$ limit gives rise to a dynamical contracted spin-flavor symmetry $SU(2N_f)$ in the baryon sector~\cite{Gervais:1983wq,Dashen:1993as}, which plays the key role in the mentioned  improvements.  A natural implementation of a combined $1/N_c$ and chiral expansions was subsequently developed~\cite{Bedaque:1995wb,FloresMendieta:2000mz}, on which this work represents a further elaboration.


%

In these proceedings, we briefly review the approach that combines these two fundamental  expansions of QCD   incorporating the constraints of the spin-flavor algebra in HBChPT.
We apply the formalism to analyze the $N$ and $\Delta$ masses and   axial charges. As an application   chiral extrapolations of lattice QCD results are studied.  The material presented here will appear  in a more extensive version elsewhere~\cite{Cordon-Goity-SU2}. 
\vspace*{-.5cm}

\section{Baryon masses and axial charges in the $\xi$-expansion}\vspace{-2mm}

This section gives a brief presentation of the combined $1/N_c$  and ChPT expansions for baryons~\footnote{We refer the reader to Ref.~\cite{Cordon-Goity-SU2} for further details.}. 
In the large $N_c$ limit, baryons have  masses~$\ord{N_c}$ \footnote{$1/N_c$ scalings: $F_\pi=\ord{\sqrt{N_c}}$, $g_A=\ord{N_c^0}$, $m_B = \ord{N_c}$ and $M_\pi = \ord{N_c^0}$.} 
and can, therefore, be  consider as static particles leading naturally to HBChPT  as description of baryon  low energy dynamics. For two light flavors,
the ground state baryon field is a multiplet of $SU(4)$ in the totally symmetric representation with $N_c$ indices, which is denoted by $\B$, and which for $N_c=3$ contains the $N$ and $\Delta$. The baryon Lagrangian is given by:
%
\bea
%
%
{\cal{L}}_{baryon} &=&\; \Bdag (i\,D_0  + g_A \,  u_i^a G^{ia}  -  \delta m_{HF}  -N_c\,c_1 M_{\pi}^2+\cdots)\B ,
\label{baryon-lagrangian}
\eea
where $D_0$ is the time component of the chiral covariant derivative,  $G^{ia}$ are $SU(4)$ spin-flavor generators~\cite{Dashen:1993as} (we note here that the usual normalization of these generators imply that the $g_A$ used here is a factor 6/5 larger than the usual one which corresponds to the physical value 1.267),  $u_\mu^a = -\frac1{F_\pi} \partial_\mu \pi^a + 2 a_\mu^a+\cdots$, $F_\pi = 92.4$ MeV is the pion decay constant, $ \pi^a$ are the pion fields and $a_\mu$ represents an   isovector axial source. The hyperfine mass splittings in the baryon multiplet  $\delta m_{HF} =\frac{C_{HF}}{N_c} \hat{S}^2$  is $\ord{1/N_c}$. The rest of the terms in the Lagrangian are built using chiral covariant derivatives, $u_\mu$, sources (among which are the quark masses) and products of spin-flavor generators. 
%
%
%
%
%
In the Lagrangian,  terms have definite chiral order but  do have different $1/N_c$ orders. For example, the Weinberg-Tomozawa term coming from the covariant derivative scales as $\sim 1/F_\pi^2$ and   is $\ord{p/N_c}$, while  for the $\pi B$ coupling, which is proportional to $\sim G^{ia}/F_\pi $,   the matrix elements of  $G^{ia}$ are $\ord{N_c}$ and   therefore the coupling is effectively $\ord{p\sqrt{N_c}}$. 
%
\vspace{-0cm}
As it was noticed long ago \cite{Cohen:1992uy}, because there is a low energy mass scale associated with $1/N_c$, namely the hyperfine splitting, the chiral and $1/N_c$ expansions do not commute: it is necessary to lock the expansions by fixing the low energy power counting of the hyperfine splitting. In strict large $N_c$ limit one could set that splitting to be $\ord{p^2}$, and in that case all quantities calculated will be analytic in $1/N_c$. In the real world with $N_c=3$ it is necessary to have a different counting scheme, namely:
~\footnote{
This particular choice is equivalent to the small-scale or $\epsilon$-expansion of Ref.~\cite{Hemmert:1997ye} and the one adopted in Ref.~\cite{Oh:1999yj}. Large $N_c$ ChPT results~\cite{Bedaque:1995wb} would be recovered by choosing $ 1/N_c \sim p^2/\Lambda_\chi^2$ and the $\delta$-expansion of Ref.~\cite{Pascalutsa:2002pi} by  choosing $ 1/N_c \sim \sqrt{p/\Lambda_\chi}$.},
\bea
\xi \sim \frac1{N_c} \sim \frac{\delta m_{HF}}{\Lambda_\chi} \sim \frac{p}{\Lambda_\chi}\, .
\eea
We call the expansion with this power counting the $\xi$-expansion.
The 1-loop calculation discussed next gives the baryon masses up to $\ord{\xi^3}$ and the axial charges up to $\ord{\xi^2}$.
\subsection{Baryon masses}\vspace*{-.6cm}
\begin{center}
\begin{figure}[h]
\centerline{\includegraphics[width=4.cm,angle=-0]{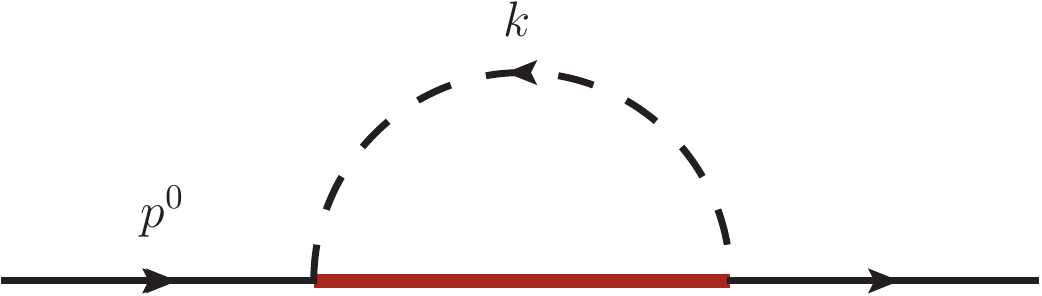}}
\caption{One-loop contribution to baryon self energy. }
\label{fig:selfnergy}
\end{figure}
\end{center}
\vspace{-1cm}
The leading 1-loop correction to the baryon self energy, diagram in Fig.~\ref{fig:selfnergy},  is given by:
\bea
 \delta\Sigma_{(1-loop)}&=&i\,\frac{g_A^2}{F_\pi^2}\;\frac{1}{d-1}\;\sum_{n} G^{ia} {\cal{P}}_n
G^{ia}\;I_{(1-loop)}(\Delta_n,p^0,M_\pi)\, , 
\eea
where the loop integral,
\bea
I_{(1-loop)}(\Delta_n,p^0,M_\pi) &=& \int \frac{d^d k}{(2 \pi)^d}\;
\frac{\vec{k}^2}{k^2-M_\pi^2+i\eps}\; \frac{1}{k^0 - \Delta_n + i\eps}
\, ,
\eea
is calculated in dimensional regularization. Here ${\cal{P}}_n$ are  the projectors onto intermediate baryon states, $\Delta_n = \delta m_n - p^0$ and $\delta m_n$ is the mass shift of the intermediate baryon.
The mass shift contains the hyperfine mass splitting plus the quark mass contribution to the baryon mass (LO $\sigma$-term), namely: $ \delta m =\delta m_{HF}+N_c\; c_1\, M_\pi^2$, which are terms of the same  order in the $\xi$-expansion. If the residual baryon mass is $\delta m_{in}$, defining $p^0 = \delta m_{in} + {\mathfrak{p}}^0$,   allows for implementation of  a convenient minimal subtraction  scheme. The UV divergent pieces are analytic in $M_\pi^2$ and in $1/N_c$, and the counter terms are organized in powers of $\xi$. In the present expansion there are non-analytic terms in $1/N_c$ in the finite parts, unlike what happens in an expansion done in the strict large $N_c$ limit. The 1-loop calculation starts at $\ord{\xi^2}$ and here terms up to $\ord{\xi^3}$ are kept, resulting in the  baryon masses:
%
\bea
m_B(S) = N_c m_0 + \frac{C_{HF}}{N_c} S(S + 1) + c_1 N_c M_\pi^2 + \delta m_{(\xi^3)}(S)  \label{eq:mass-formula},
\eea
where
\bea
\delta m_{(\xi^3)}&=&\left . \delta\Sigma^{ren}_{(1-loop)}\right \vert_{{\mathfrak{p}}^0=0}(1+\delta Z^{ren}_{(1-loop)})
\, , \hspace{1cm} 
\delta Z^{ren}_{(1-loop)}=\left . \frac{\partial  \delta\Sigma^{ren}_{(1-loop)}}{\partial {\mathfrak{p}}^0 } \right
\vert_{{\mathfrak{p}}^0\to 0},
\eea
where $ren$ indicates renormalized quantity, including all counterterms up to $\ord{\xi^3}$ (for the details  see \cite{Cordon-Goity-SU2}).  The mass shift includes a correction due to the wave function renormalization; this is so because $ \left. \delta\Sigma^{ren}_{(1-loop)}\right \vert_{{\mathfrak{p}}^0=0}$ contains terms $\ord{\xi^2}$ and  $\delta Z^{ren}_{(1-loop)}$ contains terms $\ord{\xi}$, and therefore to accuracy $\ord{\xi^3}$ that correction must be included. A few  remarks are in order: 1) unlike the HBChPT calculation with only the $N$ degree of freedom, there are several mass counterterms all involving the $\Delta-N$ mass splitting, and all counter terms are at most $\ord{N_c^0}$, and range from $\ord{\xi^2}$ to $\ord{\xi^4}$; 2) the wave function renormalization required has the same range in $\xi$, and it contains a term $\ord{N_c}$, whose presence is essential for the cancellations that take place in the axial currents' 1-loop corrections below; 4) the non-analytic contributions can be $\ord{N_c}$, as it is the case with the famous term proportional to $M_\pi^3$, however the contributions to the $N-\Delta$ mass splitting are $\ord{1/N_c}$ as they should be. \vspace*{-.2cm}

\subsection{Axial charges}\vspace*{-.3cm}
\begin{center}
\begin{figure}[h]
\centerline{
\includegraphics[width=3.5cm,angle=0]{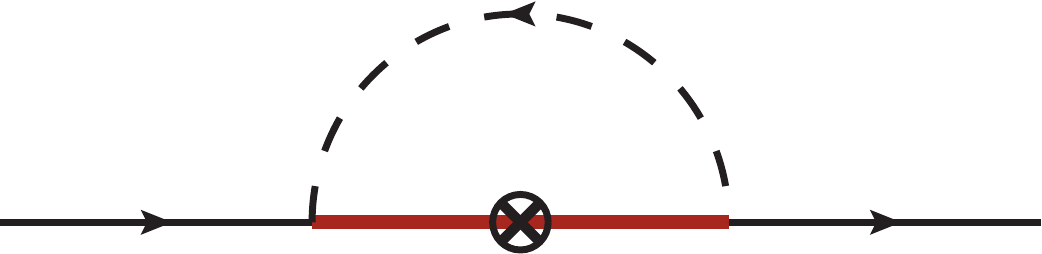}
\includegraphics[width=3.5cm,angle=0]{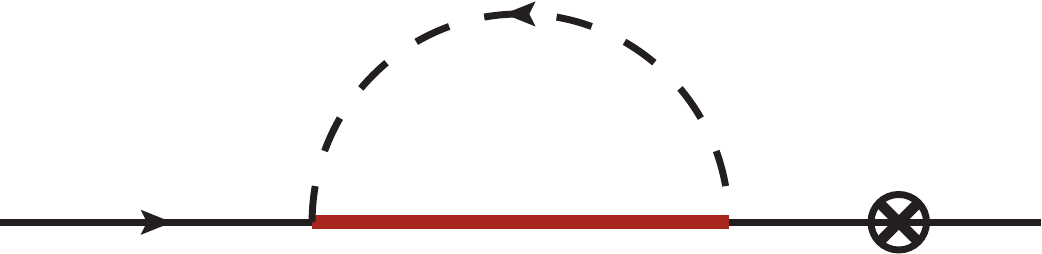}
\includegraphics[width=3.5cm,angle=0]{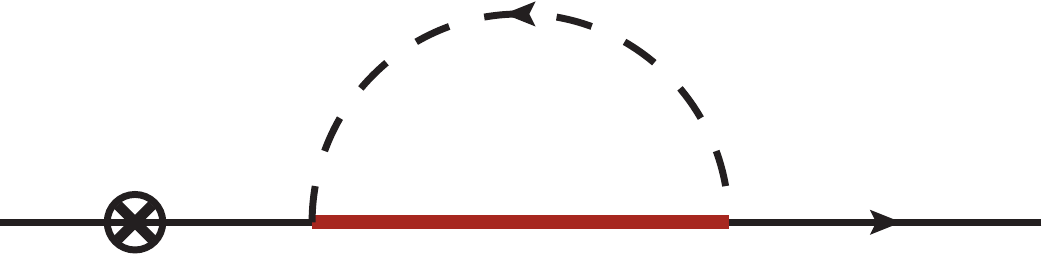}
\includegraphics[width=3.cm,angle=0]{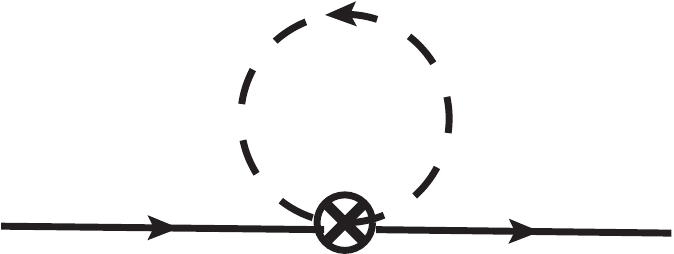}
}
\caption{Diagrams contributing to the 1-loop corrections to the axial-currents. The red lines denote the sum over different intermediate baryon states.  The crossed circle denotes   the axial-current operator.  }
\label{fig:1-loop-vertex}
\end{figure}
\end{center}

%
\vspace{-1cm}
The axial charges   to $\ord{\xi^2}$ require calculation of the 1-loop corrections to the axial current. Only the contributions with no pion pole are necessary, and they are given by
the diagrams  in Fig.~\ref{fig:1-loop-vertex}.  The resulting 1-loop contribution to the axial currents reads:
\bea
\delta A^{ia}_{(1-loop)}&=& 
\frac{{g_A}^3}{F_\pi^2}\ \frac{1}{d-1} \sum_{n,n'} G^{jb} {\cal{P}}_{n'}
G^{ia}{\cal{P}}_n G^{jb}
\frac{\left( I_{(1)}(\Delta_n,p^0,M_\pi)- I_{(1)}(\Delta_{n'},p'^0,M_\pi)\right)}
{p^0-p'^0-\delta m_n+\delta m_{n'}} \nonumber \\
&-&i\,\frac{g_A}{2} \,{( G^{ia} \delta Z_{(1-loop)} +
 \delta Z_{(1-loop)}  G^{ia}) }
+ \frac{{g_A}}{3 F_\pi^2}\,\Delta(M_\pi) G^{ia}\, ,
\eea
where $\Delta(M_\pi)$ is the pion tadpole integral.
Note that diagrams (1), (2) and (3) scale each as $\ord{N_c^{2}}$,  while the axial current must scale at most as   $\ord{N_c}$. Here is where the key cancellation of those power counting violating terms between the diagrams must occur~\cite{FloresMendieta:2000mz}. It is important to note that the required cancellation can only happen  when the contributions of the different  virtual baryon states  in the loop are included, thus showing the role of the spin-flavor symmetry in that cancellation. An important issue is the impact of that cancellation in the real world with $N_c=3$:  in fact, it is found to be very important \cite{Cordon-Goity-SU2}.
The last  diagram (4) in the figure scales as $\ord{N_c^0}$. 
Renormalizing the 1-loop calculation of the axial currents, one extracts the expression of the physical axial couplings, which will now depend on the external baryons considered, namely $NN$, $N\Delta$ or $\Delta\Delta$. They are given by evaluating the following expression in the limit of vanishing 3-momentum flowing into the current:

%
\beq
g_A^{\B\B'} = {g_A} +\frac{\langle \B'\mid\delta A^{ren~ia}_{(1-loop)}\mid \B\rangle }{ \langle \B' \mid G^{ia} \mid \B \rangle} \, .
\label{eq:gA-formula}
\eeq
Notice that  we have defined the axial charges to be $\ord{N_c^0}$ (the $\ord{N_c}$ of the current results from the operator $G^{ia}$ having $\ord{N_c}$ matrix elements).

%

Although the details will be presented elsewhere~\cite{Cordon-Goity-SU2},   a few remarks  on the axial couplings are in order.  In the strict large $N_c$ limit: 1)  the corrections  to the axial charges  start at   $\ord{p^0 /N_c}$ and  $\ord{p^2 N_c^{ 0}}$,  and  2) their spin-flavor symmetry breaking  starts with a term in the Lagrangian   $\ord{p^0 /N_c^2}$, and the 1-loop corrections also give  $\ord{p^0/N_c^2}$ effects, while the chiral logs appear  first at $\ord{p^2/N_c^3}$. In  the $\xi$-expansion:  3)  in the present expansion,  the corrections to  axial charges and their spin-flavor breaking start at  $\ord{\xi}$.

\section{Application to lattice QCD results}\vspace*{-.5cm}
\begin{center}
\begin{figure}[h]
\centerline{\includegraphics[width=6.5cm,angle=-0]{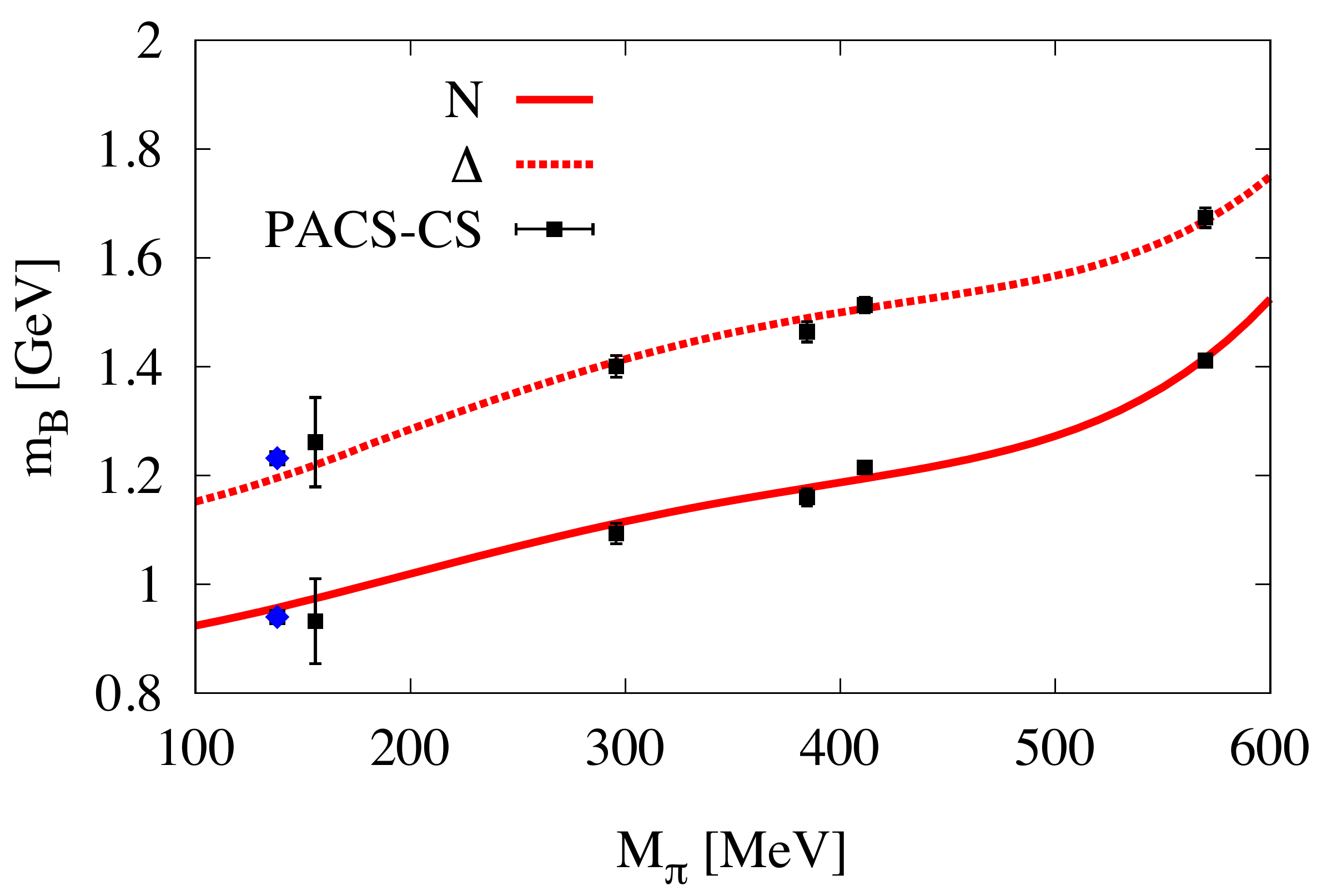}
\hspace{1cm}
 \includegraphics[width=6.5cm,angle=0]{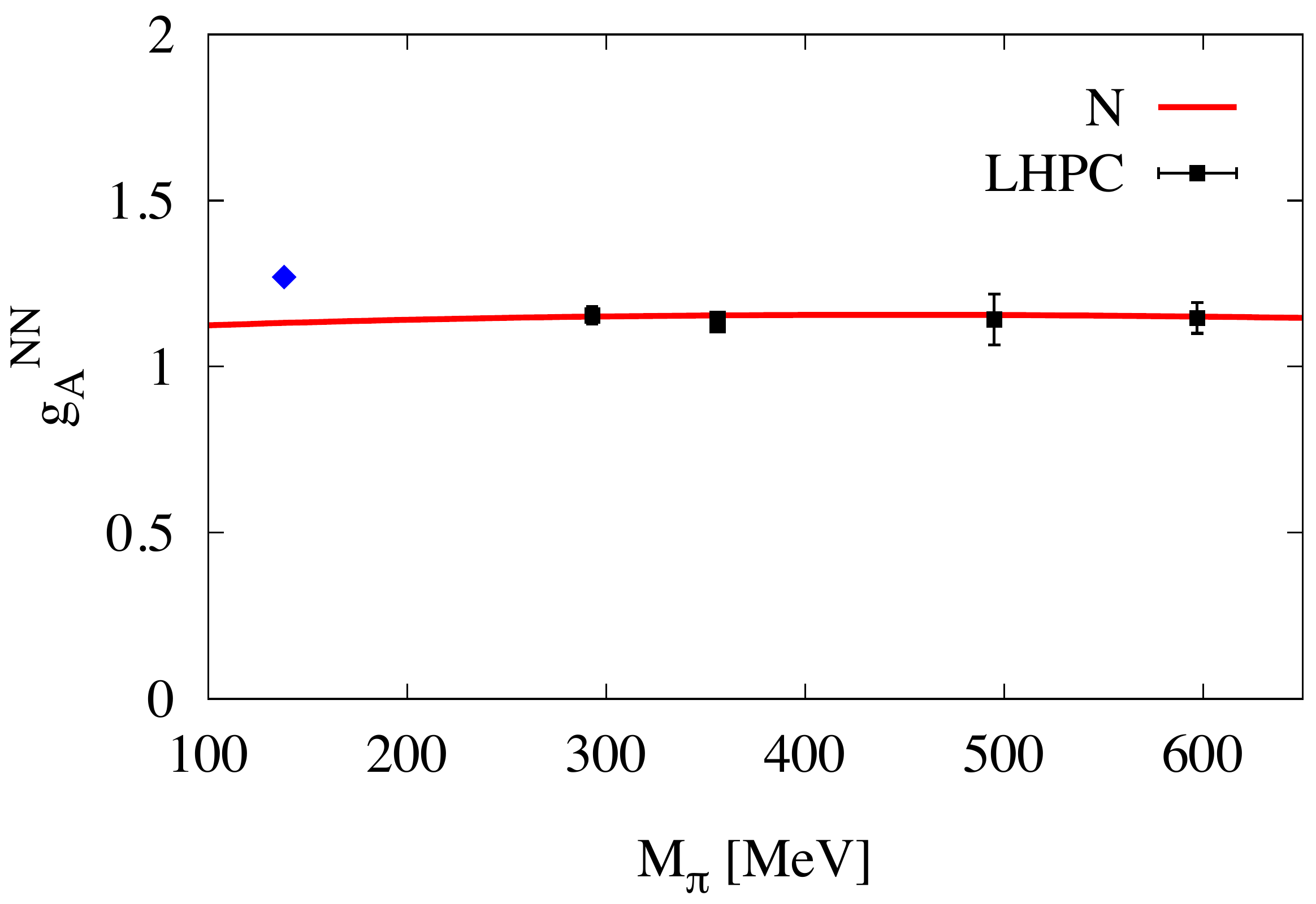}
}
\caption{Combined fit 
to lattice QCD results. Nucleon and $\Delta$ masses (left panel) are fitted to PACS-CS~\cite{Aoki:2008sm} results and the $N$ axial charge (right panel) to LHPC~\cite{Bratt:2010jn} results.  The $g^{NN}_A$ plotted here is the one with the standard normalization. }
\label{fig:lattice-fits}
\end{figure}
\end{center}

%
%
%
%
\vspace{-1cm}
Lattice QCD calculations of the non-strange ground state baryon masses (both of $N$ and $\Delta$ baryons) have opened the possibility of determining the quark mass dependencies, and similarly for the axial coupling of the nucleon. This represents a very fruitful ground of applications for ChPT, allowing in particular for a   study of the convergence of the low energy expansion. As an illustration,  we present here an analysis to $\ord{\xi^3}$ of the $N$ and $\Delta$ masses  and to $\ord{\xi^2}$ of   $g_A^{NN}$ (1-loop calculations) .
%
We perform a combined fit to the $N$ and $\Delta$ masses obtained by the PACS-CS~\cite{Aoki:2008sm} collaboration and to   $g_A^{NN}$ from the LHPC~\cite{Bratt:2010jn} collaboration. It turns out that various low  energy constants (LECs) require knowledge of results at different $N_c$, and thus they combine with existing ones at lower order for a fit at fixed $N_c=3$.  Since the fits only involve   $g_A^{NN}$, not all LECs affecting the axial currents can be determined. Finally, among the fitted LECs one finds some important correlations, which permit to set some of them to zero. Basically the fit   involves the following LECs: $m_0,~g_A, ~C_{HF},~c_1$  which appear in the terms displayed in Eq. (2.1), and in addition three higher order LECs associated to the pion mass dependence to the baryon hyperfine mass splitting,  the pion mass dependence of the baryon wave function renormalization, and   the pion mass dependence of $g_A$. %
Our results are plotted in Fig~\ref{fig:lattice-fits}. A detailed  analysis of the fit presented here as well as fits of results from other lattice QCD calculations will be presented in Ref.~\cite{Cordon-Goity-SU2}.

Remarks on the fits: 1) all fitted parameters  (LECs) are of natural magnitude when the renormalization scale is taken to be $\mu\sim m_\rho$; 2) parameters appearing at lower orders, namely $m_0,~g_A, ~C_{HF}$ remain stable at higher orders, an exception being  $c_1$ which changes by more than the estimated 30\% when increasing the order of the fit by one unit in $\xi$; 3) the cancellations of large contributions from separate loop diagrams to $g_A$ are very pronounced and the almost flat behavior  as function of $M_\pi$ obtained in lattice QCD  is naturally explained; 4) the physical $g_A^{NN}$ cannot be fitted along with the lattice results, instead the lattice results and the expansion to $\ord{\xi^2}$ of $g_A^{NN}$ extrapolate to a 12\% smaller value than the physical one; 5) a fit  restricted only to masses  gives too small a value for $g_A$, thus  realistic values for the axial charges require obviously the combined fit; 6) relating  our work outlined here to recent  ChPT  analyses of lattice results \cite{WalkerLoud:2011aa}  will be pursued in Ref.~\cite{Cordon-Goity-SU2}.\vspace{-3mm}
 
\section{Comments}
The presence of an  approximate dynamical spin-flavor symmetry in baryons ($SU(4)$ and $SU(6)$) has been known for ages, and it is a result of the large $N_c$ limit of QCD. The breaking of spin-flavor symmetry can be studied in a rigorous way by performing the $1/N_c$ expansion at baryon level. In addition, it can be combined with ChPT   leading to significant improvements in the low energy description of baryons. The exciting progress in lattice QCD is now opening the opportunity for testing   the different versions of the baryon low energy effective theories. At the current stage,  a strong indication supporting the version of ChPT which includes the $1/N_c$ expansion is emerging, as previous analyses and  the one described here indicate. Of particular interest to further advance the study presented here are    lattice QCD calculations of $g_A^{N\Delta}$ and  $g_A^{\Delta\Delta}$, on which  pioneering work has already been done~\cite{Dina}. 
 In due time we expect that also baryon observables will be calculable  in lattice QCD at larger values of $N_c$  (this has been recently initiated in the work of Ref. \cite{DeGrand}), which will help us draw more definite conclusions about the practical virtues of the $1/N_c$ expansion.  

\end{document}